\documentclass{mem}
\usepackage{natbib}\usepackage{txfonts}\usepackage{balance}
\usepackage{graphicx}
\usepackage[a4paper,breaklinks,dvipdfm]{hyperref}
\idline{84}{1}
\usepackage{txfonts} 

\begin{document}

\title{A scientific case for future X-ray Astronomy: Galaxy Clusters at high redshifts}

\subtitle{}

\author{P. \,Tozzi}

\offprints{P. Tozzi}
 
\institute{
INAF -
Osservatorio Astrofisico di Arcetri, Largo E. Fermi,
I-50122 Firenze, Italy\\
\email{ptozzi@arcetri.astro.it}
}

\authorrunning{Tozzi}

\titlerunning{Galaxy Clusters at high-z}

\abstract{Clusters of galaxies at high redshift ($z>1$) are vitally important to understand 
the evolution of the large scale structure of the Universe, the processes 
shaping galaxy populations and the cycle of the cosmic baryons, and to constrain
cosmological parameters.  After 
13 years of operation of the Chandra and XMM-Newton satellites, the 
discovery and characterization of distant X--ray clusters is proceeding at a 
slow pace, due to the low solid angle covered so far, and the time-expensive 
observations needed to physically characterize their intracluster medium (ICM).  
At present, we know that at $z>1$ many massive clusters are fully virialized, 
their ICM is already enriched with metals, strong cool cores are already in place, and 
significant star formation is ongoing in their most massive galaxies, at least at $z>1.4$.  Clearly, 
the assembly of a large and well characterized sample of high-z X-ray clusters is a major goal
for the future.  We argue that the only means to achieve this is a 
survey-optimized X-ray mission capable of offering large solid angle, high 
sensitivity, good spectral coverage, low background and angular resolution
as good as 5 arcsec.

}
\maketitle{}

\section{Introduction}

X-ray observations of clusters of galaxies over a significant range
of redshifts have been used to investigate the chemical and
thermodynamical evolution of the X-ray emitting intracluster medium (ICM)
and to constrain the cosmological parameters and the spectrum of the
primordial density fluctuations.
In this respect, X-ray surveys of clusters of galaxies represent a key tool
for cosmology and the physics of large scale structure.  
The compilation of more and larger X-ray selected samples with well-defined completeness criteria
is one of the critical requirements of present-day cosmology.  

There is a remarkable lack of recent wide-area X-ray surveys
suitable to this scope.  Most of the existing cluster samples are still based 
on ROSAT data.  The most recent constraints on cosmological parameters 
from X-ray clusters are based on the Chandra follow-up of 400 deg$^2$ 
ROSAT serendipitous survey and of the All-Sky Survey \citep{2009Vikhlinin,2010Mantz}.  
At present, there are no cosmological constraints based on the mass function of 
clusters selected from Chandra or XMM-Newton data.  There is, however, a significant effort 
devoted to new X-ray surveys based on  data from currently operating X-ray satellites, 
realized thanks to compilations of serendipitous 
medium and deep extragalactic fields not associated with previously known X-ray clusters, 
plus a few dedicated, contiguous observations.  A table summarizing the properties of 
post-ROSAT cluster surveys is presented in Tundo et al. (2012).

\begin{figure}
\centering \includegraphics[width=\columnwidth]{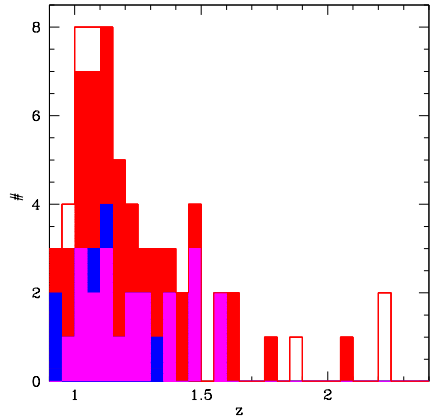}
\caption{\footnotesize
Redshift histrogram distribution of clusters at $z>1$ reported in the literature as of
September 2012.  Colors corresponds
to different selection methods: red=IR; magenta=X-ray; blue=SZ; white=other.}
\label{highzclusters}
\end{figure}

One of the most important goals of these surveys is to find massive
clusters at high $z$.   If we focus on the most distant clusters
reported in the literature, irrespective of the
selection band, we find that only nine have been
spectroscopically confirmed at $z\ge 1.5$ to date, and only a few  of
them have estimated masses in excess of $10^{14}$ $M_\odot$.
The redshift histogram distribution of clusters and cluster candidates at $z>1$ (less than 70 overall) is shown in
Figure \ref{highzclusters}. While the redshift limit for X-ray selected clusters is $z\sim 1.57$
\citep{2011Fassbender}, extended X-ray emission has been detected in optical and
IR selected clusters out to $z=1.75$ \citep{2012Stanford}, $z=2.07$
\citep{2011Gobat}, and $z\sim 2.2$  for an extreme candidate \citep{2012Andreon}.   
However, many other cluster candidates identified in the IR  
are undetected, suggesting uncomplete virialization. 

Present-day X-ray facilities offer a limited discovery space mostly because of the small solid
angle covered so far.  Although the exploitation
of the Chandra and XMM archives is still far from being concluded,
we can predict, on the basis of what has been obtained to date, 
that in the future the number of clusters firmly detected in X-ray at $z>1$
will grow up to $\sim 100$ in the most optimistic case.  Note that this number
applies only to the detection of the cluster X-ray emission.  
In fact, the number of high-z X-ray cluster with a robust physical 
characterization will be at best a factor of 3 lower.
The reason is that the only means to determine the virialization status of
distant clusters, to measure their mass and characterize the
thermodynamical and chemical properties of  their ICM, is to perform
time-expensive observations with Chandra (of the order of few $\times 100$ ks)
to collect more than 1000 net photons in the 0.5-7 keV band in high resolution images to remove the
effect of contaminating AGN emission and of central cool cores. 

This situation is not expected to improve on the basis of currently planned missions.
Looking at the near future, the upcoming
mission eROSITA will not be efficient in finding extended sources 
at fluxes below few $\times 10^{-14}$ erg cm$^{-2}$ s$^{-1}$, where most of the distant
clusters are.  Finally, although X-ray observations still dominate
this field, Sunyev-Zeldovich (SZ) cluster surveys are rapidly gaining momentum,  thanks to 
the South Pole Telescope Survey \citep{spt2012} and the Atacama
Cosmology Project \citep{act2011}, which will be soon complemented by
a shallower, but all-sky survey by the Planck satellite.  At present, the most distant confirmed
 SZ selected cluster is reported at $z\sim 1.32$ \citep{2012Stalder}.  The X-ray data, however, are strongly
 complementary to SZ, and a complete characterization of a galaxy cluster, particularly
 a distant one, must rely on both X-ray and SZ data.  In order to keep the crucial role
 of X-ray astronomy in this field, a wide and deep coverage of the X-ray sky is
needed.  This can be achieved only with a dedicated mission optimized for surveys, 
as we argue in the following sections.

\section{The Swift X-ray cluster survey}

\begin{figure}
\centering \includegraphics[width=\columnwidth]{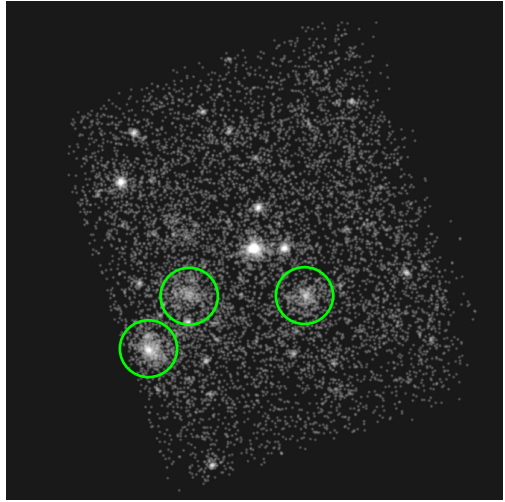}
\caption{\footnotesize
XRT smoothed image of the GRB050505 field in the soft
  ($0.5-2.0$ keV) band; the exposure time is $\sim$170 ks, and the
  image size is $\sim$24$\times$24 arcmin. The three clusters in the
  field are highlighted by the circles.  The bright source at
  the center is the GRB (from Tundo et al. (2012).}
\label{xrt_ima}
\end{figure}

A constant angular resolution across the field of view (FOV) and a low background are two key
properties that a future X-ray survey must retain in order to efficiently
detect and characterize faint extended sources, such as distant X-ray clusters.
The Swift X-ray Cluster Survey  (SXCS), built upon the rich set of archival data of
the X-ray Telescope (XRT) onboard the Swift satellite
\citep{2005Burrows}, is an excellent example of what can 
be done thanks to these two key properties.  
The point spread function (PSF) of XRT is characterized by a half energy width
(HEW) of $\sim 18$ arcsec at 1.5 keV, with negligible degradation with
the off-axis angle on the entire FOV of 24 arcmin \citep{2007Moretti}.   
The SXCS catalog presented in Tundo et al. (2012) is based on the 336 GRB
fields with galactic latitude $|b|>$20$^\circ$ present in the XRT
archive as of April 2010.  The extended source candidates are identified thanks to a simple
growth-curve analysis applied to the soft band ($0.5$-$2$ keV) images (see Figure \ref{xrt_ima}).
The algorithm is validated via extensive simulations.  One of the most relevant aspects is the low level and the high reproducibility of the background not associated with astronomical sources.  Due to the
low orbit and short focal length, this background component per solid angle, normalized by
the effective area, is a factor $\sim 7$ lower than in Chandra  in the
0.5-7.0 keV energy band \citep{2012Moretti}.

The final catalog consists of 72 X-ray  group and
cluster candidates (mostly at moderate redshifts) with a negligible contamination, a well defined
selection function and a robustly estimated completeness, spanning two
orders of magnitude in flux down to the limit of $\sim
10^{-14}$ erg s$^{-1}$ cm$^{-2}$.  The corresponding logN-logS is in
very good agreement with previous deep surveys.
Thanks to a cross correlation with the NED database,
we are able to  associate 24 optical redshifts (spectroscopic or
photometric) published in the literature with our cluster candidates.
Overall, 46 sources in our catalog are new detections, both as X-ray
sources and as clusters of galaxies.   We also
measured the redshift for about half of the sources 
thanks to the identification of the redshifted $K_\alpha$ Fe line in the spectral analysis,
following the procedure described in Yu et al. (2011).  At present, the largest measured redshift
in the SXCS sample is $z = 1$, while 
many redshift measurements are still lacking for the faintest part of the sample.   This 
shows that despite its low collecting area (about one fifth of Chandra at 1 keV) XRT 
is unexpectedly providing a well characterized X-ray survey reaching 
the realm of distant galaxy clusters.  To summarize, although XRT is not able to offer 
the large solid angle and the sensitivity needed to find and characterize a consistent number of high-z clusters, 
we argue that flat PSF across the FOV
and low background are two fundamental properties for future  missions aiming at bringing the X-ray
sky to the same depth and richness of the optical and IR sky in the
next decade. 

\section{The WARPJ1415 field}

To obtain a robust characterization of distant X-ray clusters, including the measure of
the ICM temperature, its iron abundance, and hence the redshift from the 
detection of the Fe emission line complex at 6.7-6.9 keV, a relatively high effective area in the
energy range 2-7 keV is also required.  In addition, as opposed to the mere
identification of an extended source, which requires about 20 counts,
a meaningful, single-temperature spectral analysis requires a 1000-1500 net counts (in the 0.5-7 keV
band).   On the other hand, about $\sim 10000$ net counts are required if we aim at a spatially resolved
spectral analysis, which allows a much better mass
determination, a study of the iron distribution, and a detailed
characterization of the central regions (within $r_{500}$).
These science goals can be achieved at any redshift  only if the angular resolution is equal or
better than 5 arcsec HEW \citep[see][]{2011Santos}.  

A remarkable case to illustrate both regimes is the 
WARPJ1415 field which has been observed with Chandra ACIS-S
for 280 ks (plus 90 ks with ACIS-I) with the aim of studying
the  properties of the cool core of WARPJ1415 at $z= 1.03$
\citep[see][]{2012Santos}.  The sharp and deep Chandra data allowed us to measure a
very low central cooling time $t_{cool} \leq 0.23$ Gyr in the inner 20 kpc, with a surprisingly
pronounced metallicity peak, $Z_{Fe}= 3.60^{+1.50}_{-0.85} Z_\odot$.  The cluster total X-ray
mass, under the assumption of hydrostatic equilibrium, is accurately measured as
$M_{200}$=3.0$_{-0.4}^{+0.6}\times$10$^{14}$M$_\odot$.  Using VLA
data we detected a radio source coincident with the brightest central galaxy (BCG) with a
faint, one-sided structure extended for 80 kpc in the
north-west direction, where a significant lack of X-ray emission was
found.  Thanks to this study, we are able to confirm that a
feedback mechanism powered by AGN in the radio mode, as observed in 
local clusters, is at work also in the core of WARPJ1415.  In addition, the prominent iron
peak indicates that the metal enrichment mechanisms by type Ia
supernovae and star formation in the BCG happened on a short
timescale (given the large lookback time of 7.8 Gyr), and that the
transport processes that drive away the metals to the outskirts
were not efficient to smear out the iron excess.
Overall, these observations enabled the most detailed
analysis of the ICM of a $z \sim 1$ cluster to date and highlight
the importance of deep, high-resolution data to adequately
characterize distant clusters.  With current X-ray
facilities, we expect that only few cases can be studied with an
accuracy comparable to that of WARPJ1415.

\begin{figure}
\centering \includegraphics[width=\columnwidth]{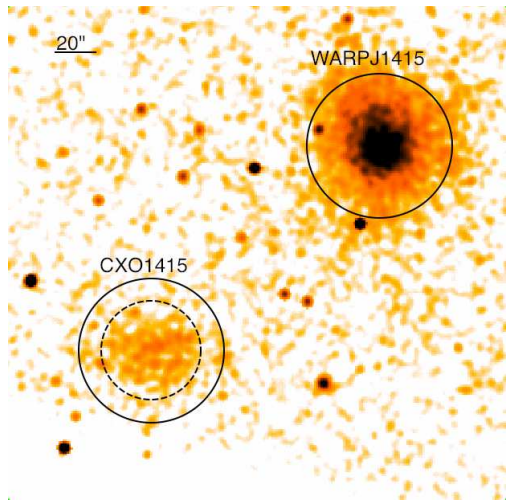}
\caption{\footnotesize
ACIS-S+I image of the WARPJ1415 field.  The bright cluster WARPJ1415 at $z=1$ is
  visible in the upper right, whereas CXO1415 at $z\sim 1.5$ lies at $\sim 2 \arcmin$
  in the south-west direction.  The solid circles have a radius of
  $35 \arcsec$ (corresponding to $300$ kpc at $z=1.46$), whereas the
  dashed circle with $r =24 \arcsec$ refers to the aperture of maximum
  S/N.  The image has a size of $ 4
  \arcmin \times 4 \arcmin$ \citep[see][]{2012Santos}.}
\label{xray}
\end{figure}

As an unexpected bonus, CXO J1415 was serendipitously detected in the same
observation \citep[see][]{2013Tozzi}.  The new system, located at a distance of $\sim 2$ arcmin
in the south-west direction of WARPJ1415, was immediately detected by
visual inspection as an extended X-ray source in the ACIS-S image
(see Figure \ref{xray}) with a $S/N\sim 11$ within a
  radius of 24 arcsec.  The soft-band flux within $r=24\arcsec$ is equal to
  $S_{0.5-2.0 keV}=(6.5 \pm 0.3) \times 10^{-15}$ erg s$^{-1}$
  cm$^{-2}$.  We measure a redshift of $z=1.46\pm 0.025$ by
  identifying the Fe $K_\alpha$ complex line at a $\sim 2.8
  \sigma$ confidence level.  This shows that the measure of redshift
  through X-ray spectral analysis is possible up to the highest
  redshift where X-ray clusters are currently detected.
This makes CXO1415 the highest-z cluster discovered with Chandra so far. 
The spectral fit with a {\tt mekal} model gives $kT=5.8^{+1.2}_{-1.0}$ keV, for a total mass of $M(r<300 {\tt kpc}) =
  1.38_{-0.28}^{+0.33} \times 10 ^{14}M_\odot$, while the ICM mass is
  $M_{ICM} (r<300 {\tt kpc}) = 1.09_{-0.20}^{+0.30}\times 10^{13}M_\odot$,
  resulting in a ICM mass fraction of $\sim 13$\%.  The color-magnitude diagram of 
  cluster member candidates shows both a red sequence and a significant fraction of blue,
  irregular-morphology galaxies.  Finally, when compared with the expectations 
  for a $\Lambda$CDM universe, CXO1415 appears to
  be a typical cluster at $z\sim 1.5$ for a WMAP cosmology
  \citep{komatsu11}; however, the redshift and the total mass of
  CX01415 place it among the sample of massive, distant galaxy clusters
  which may be used to accurately test the standard $\Lambda$CDM model once a
  sizeable sample of high-z clusters will be assembled.

\section{Conclusions}

After 13 years of operations of the Chandra and XMM-satellites, we have
a robust vision of cluster properties at $1<z<1.5$, and a glimpse  of it at $z>1.5$. Now we know that
not only massive X-ray clusters are
abundant at $z>1$, but also that at $z>1.5$ we keep finding clusters whose ICM is already virialized, strongly enriched
with metals, and whose color-magnitude relation is already partially established.  
Without any doubt, high-z, X-ray clusters represent a scientific case which can provide critical insights
into the physics of cosmic structure formation, galaxy evolution, the life cycle of the cosmic baryons, and
provide constraints on cosmological parameters.  

At present, the investigation of distant X-ray clusters is proceeding slowly.  
High-z clusters are rare, faint objects with a small extension ($\leq 1$ arcmin), and require deep and wide-angle observations, with a good angular resolutions, in order to be found.  Moreover, a good spectral coverage up to 7 keV and a large collecting area are needed to physically characterize them.  In this perspective, the future looks grim.
The planned eROSITA satellite \citep{2010Predehl} will be confusion limited in the flux regime where the
majority of the high-$z$ clusters are expected.  In addition, its low
hard-band sensitivity hampers a robust measure of the
temperature for hot clusters, limiting cosmological studies.
The only means to address this scientific case is a survey-optimized mission
such as the proposed Wide Field X-ray Telescope \citep[WFXT][]{2010Murray}. 
The WFXT, thanks to its good angular resolution and a
constant image quality across a 1 deg$^2$ FOV, coupled with a large
effective area, will provide a direct measurement of
temperatures, density profiles and redshifts for about 1000
X-ray clusters at $z>1$, without the need of 
time-prohibitive spectroscopic follow-up programs.  This 
will improve by almost two orders of magnitude any 
well-characterized cluster sample that we can possibly assemble 
using the entire wealth of data from present and planned X-ray facilities.
We conclude that a mission such as WFXT is able to fully address the 
scientific case of high-z clusters, among many others, as well as to match in
depth, survey volume and angular resolution surveys at other
wavelengths which are planned for the next decade.

\begin{acknowledgements}

We acknowledge support from contract ASI-INAF
I/088/06/0 and ASI-INAF I/009/10/0.  We thank L. Hunt, R. Gilli, 
M. Paolillo and P. Rosati for helpful comments.

\end{acknowledgements}

\bibliographystyle{aa}

\end{document}